\documentclass[twocolumn,showpacs,preprintnumbers,showkeys,amsmath,amssymb,aps,pra,10pt]{revtex4-1}

\usepackage{graphicx}
\usepackage{dcolumn}
\usepackage{bm}

\usepackage[T1]{fontenc} 
\newcommand{\wyroz}[1]{#1}

\begin{document}

\preprint{Submitted to: JOURNAL OF SUPERCONDUCTIVITY AND NOVEL MAGNETISM}

\title{Interplay and competition between superconductivity and charge orderings\\
in the zero-bandwidth limit of the extended Hubbard model\\ with pair hopping and on-site attraction}%

\author{Konrad Kapcia}%
    \email{e-mail: kakonrad@amu.edu.pl}
\affiliation{Electron States of Solids Division, Faculty of Physics, Adam Mickiewicz University in Pozna\'n, Umultowska 85, PL-61-614 Pozna\'n, Poland, EU
}%

\date{March 6, 2013}

\begin{abstract}
We present studies of  an effective model which is a simple generalization of the standard model of a local pair superconductor with on-site pairing (i.e., the model of hard core bosons on a lattice) to the case of finite pair binding energy.  The tight binding Hamiltonian consists of (i) the effective on-site interaction $U$, (ii) the intersite density-density interactions $W$ between nearest-neighbours, and (iii) the intersite charge exchange term $I$, determining the hopping of electron pairs between nearest-neighbour sites.
In the analysis of the phase diagrams and thermodynamic properties of this model we treat the intersite interactions within the mean-field approximation.
Our investigations of the $U<0$ and $W>0$ case show that, depending on the values of interaction parameters, the system can exhibit three homogeneous phases: superconducting (SS), charge-ordered (CO) and nonordered (NO)
as well as the phase separated SS--CO state.
\end{abstract}

\pacs{
71.10.Fd --- Lattice fermion models (Hubbard model, etc.),
74.20.-z --- Theories and models of superconducting state,
64.75.Gh --- Phase separation and segregation in model systems (hard spheres, Lennard-Jones, etc.),
71.10.Hf --- Non-Fermi-liquid ground states, electron phase diagrams and phase transitions in model systems,
71.45.Lr --- Charge-density-wave systems}
\keywords{extended Hubbard model, phase separation, superconductivity, charge orderings, local pairing, phase diagrams}
\maketitle

\section{Introduction}
\label{intro}

The interplay and competition between superconductivity and charge orderings is currently under intense investigations (among others in high temperature superconductors such as cuprates, barium bismuthates, fullerenes and several other nonconventional superconducting materials, e.g., the Chevrel phases) \cite{MRR1990}. They belong to a unique group of extreme type II superconductors and generally exhibit low carrier density, a~small value of Fermi energy ($E_F\leq0.1 \div 0.3$ eV) and a short coherence length $\xi_0$ ($\xi_0k_F\approx1\div10$). These general features are consistent with short-range, almost unretarded effective interactions response for local pairing.

In this report we will concentrate on the intriguing problem of phase separation and the competition between superconductivity and charge orderings. The model Hamiltonian considered has the following form:
\begin{eqnarray}\label{row:ham}
\hat{H} & = & U\sum_{i}{\hat{n}_{i\uparrow}\hat{n}_{i\downarrow}} - \mu\sum_i\hat{n}_i\\
& - & 2I\sum_{\langle i,j\rangle}{\hat{\rho}_i^+\hat{\rho}_j^-} +\frac{W}{2}\sum_{\langle i,j \rangle} \hat{n}_i\hat{n}_j, \nonumber
\end{eqnarray}
where  \mbox{$\hat{n}_{i}=\sum_{\sigma}{\hat{n}_{i\sigma}}$}, \mbox{$\hat{n}_{i\sigma}=\hat{c}^{+}_{i\sigma}\hat{c}_{i\sigma}$}, \mbox{$\hat{\rho}^+_i=(\hat{\rho}^-_i)^\dag=\hat{c}^+_{i\uparrow}\hat{c}^+_{i\downarrow}$}.
$\hat{c}^{+}_{i\sigma}$ ($\hat{c}_{i\sigma}$) denotes the creation (annihilation) operator of an electron with spin \mbox{$\sigma=\uparrow,\downarrow$} at the site $i$,
which satisfy canonical anticommutation relations:
\begin{equation}
\{ \hat{c}_{i\sigma}, \hat{c}^+_{j\sigma'}\} = \delta_{ij}\delta_{\sigma\sigma'},
\{ \hat{c}_{i\sigma}, \hat{c}_{j\sigma'}\} = \{ \hat{c}^+_{i\sigma}, \hat{c}^+_{j\sigma'}\} = 0,
\end{equation}
where $\delta_{ij}$ is the Kronecker delta.
\mbox{$\sum_{\langle i,j\rangle}$} indicates the sum over nearest-neighbour sites $i$ and $j$ independently.
$z$ will denote the number of nearest-neighbours.
$U$, $I$, and $W$  are the interactions parameters, $I_0=zI$, $W_0=zW$.
$\mu$ is the chemical potential, connected with the concentration
of electrons by the formula:
%
$n = (1/N)\sum_{i}{\left\langle \hat{n}_{i} \right\rangle}$,
%
with \mbox{$0\leq n \leq 2$} and $N$ is the total number of lattice sites.

The interactions $U$, $I$, and $W$ will be treated as the effective ones and will be assumed to include all the possible contributions and renormalizations like those coming from the strong electron-phonon coupling or from the coupling between electrons and other electronic subsystems in solid or chemical complexes \cite{MRR1990}. In such a general case, arbitrary values and signs of $U$, $I$, and $W$ are important to consider.
In the absence of the external field conjugated with the SS order parameter ($\Delta=(1/N)\sum_i\langle \hat{\rho}^-_i\rangle$)  there is a symmetry between \mbox{$I>0$} ($s$-pairing) and \mbox{$I<0$} ($\eta$-pairing, $\eta$S, $\Delta_{Q} = (1/N)\sum_i{\exp{(i\vec{Q}\cdot\vec{R}_i)}\langle \hat{\rho}^-_i\rangle} $, $\vec{Q}$ is half of the smallest reciprocal lattice vector) cases for model (\ref{row:ham}), thus we restrict ourselves only to the \mbox{$I>0$} case. The CO parameter is defined as $n_Q=(1/N)\sum_i{\exp{(i\vec{Q}\cdot\vec{R}_i)}\langle \hat{n}_i\rangle}$.

\begin{figure*}
    \centering
    \includegraphics[width=1.0\textwidth]{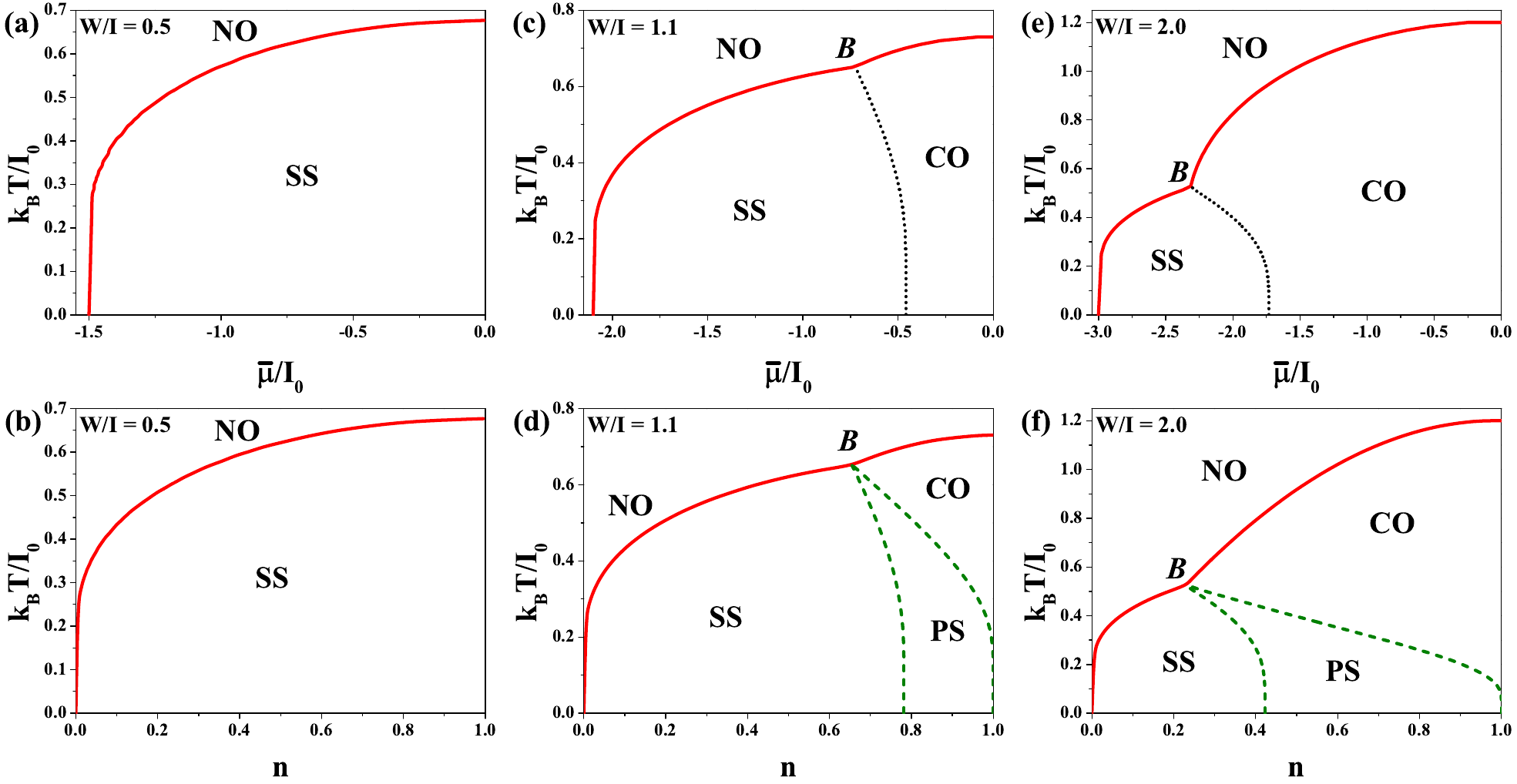}
    \caption{$k_BT/I_0$~vs.~$\bar{\mu}/I_0$ phase diagrams (upper row) and corresponding $k_BT/I_0$~vs.~$n$ diagrams (lower row)  for \mbox{$U/I_0=-1.0$} and different values of \mbox{$W/I=0.5,\ 1.1,\ 2.0$} (as labelled). Dotted, solid, and dashed lines indicate first-order, second-order, and ``third-order'' boundaries, respectively. $\mathbf{B}$ denotes bicritical points.}
    \label{rys:PSU0}
\end{figure*}

We have performed extensive study of the phase diagrams (PDs) of model (\ref{row:ham}) \cite{KR0000}. In this report we investigate the properties of model (\ref{row:ham}) for the case of on-site attraction ($U<0$, \emph{local pair} or \emph{Bose condensation} limit \wyroz{\cite{RP1993,R1994,KRM2012,KR2013,BR1996,RB1999}}) and intersite repulsion ($W>0$).
In the analysis of the model at \mbox{$T\geq 0$}, we have adopted a variational approach (VA) which treats the on-site interaction $U$ exactly and the intersite interactions $W$ and $I$ within the mean-field approximation (MFA).
The PDs of model (\ref{row:ham}) have been investigated until now for the special cases: $W=0$ \wyroz{\cite{RP1993,R1994,KRM2012,KR2013,BR1996,RB1999,K2012,B1973,HB1977,WA1987}} and $I=0$ \wyroz{\cite{R1973,MRC1984,MM2008,MM2010,KKR2010,KR2011,KR2011a,KR2012,KKR2012}} only. Some ground state results for $W\neq0$ and $I\neq0$ have been also obtained \cite{KRM2012,RP1996,PR1996}.

Within the VA the intersite interactions are decoupled within the MFA, which allows us to calculate the averages $n$, $n_Q$, $\Delta$, and $\Delta_Q$. It gives a set of four self-consistent equations (for homogeneous phases).
The definitions of homogeneous phases with the values of order parameters are as follows: (i) SS --  $n_Q=0$, $\Delta\neq0$, $\Delta_Q=0$; (ii)
CO -- $n_Q\neq0$, $\Delta=0$, $\Delta_Q=0$; (iii) M -- $n_Q\neq0$, $\Delta\neq0$, $\Delta_Q\neq0$; (iv) NO -- $n_Q=0$, $\Delta=0$, $\Delta_Q=0$.
It is important to find a solution corresponding to the lowest energy.

Phase separation (PS) is a state in which two domains with different electron concentration: $n_+$ and $n_-$ exist in the system
(coexistence of two homogeneous phases). The free energies of the PS states are calculated in a~standard way, using Maxwell's construction (e.g., Refs.~\cite{KRM2012,KR2011,KR2011a,B2004}). In model (\ref{row:ham}) for the  range of parameters considered in this paper only one PS state can occur, which is coexistence of the SS and CO phases.

In the report, we have used the following convention. A~second- (first-)order transition is a~transition between homogeneous phases  with a~(dis-)continuous change of the order parameter at the transition temperature. A~transition between homogeneous phase and PS state is symbolically named as a~``third-order'' transition~\wyroz{\cite{KRM2012,KR2013,KKR2010,KR2011,KR2011a,KR2012,KKR2012}}.

\section{Results in the local pairing limit}

\subsection{Phase diagrams}

One should noticed that PDs obtained are symmetric with respect to half-filling (\mbox{$n=1$}) because of the particle-hole symmetry of Hamiltonian (\ref{row:ham}), so the PDs will be presented only in the range \mbox{$\bar{\mu}=\mu-U/2-W_0 \leq 0$} and \mbox{$0\leq n\leq 1$}.

For any \mbox{$U\leq0$} and fixed $W>0$, the PDs are qualitatively similar, all (first-order, second-order and ``third-order'') transition temperatures decrease monotonically with increasing $U$ and in the VA for \mbox{$U=0$} the transition temperatures account for a~half of those in the limit \mbox{$U\rightarrow-\infty$}, what can be symbolically written as
\begin{equation}
k_B T_c(U\rightarrow-\infty) = 2 k_B T_c (U=0),
\end{equation}
where $T_c$ denotes the transition  temperature (which can be SS-NO, SS-CO, CO-NO, PS-CO or PS-SS).

Notice that in the \mbox{$U\rightarrow-\infty$}  limit model (\ref{row:ham}) is equivalent with \wyroz{the} hard-core boson model on a~lattice \cite{GR1998,GMR1998,MR1992,MRK1995,STTD2002}. Moreover in that limit model (\ref{row:ham}) can be derived as effective Hamiltonian at the strong-coupling limit of the extended Hubbard model by the degenerate perturbation theory~\cite{RP1996,PR1996,RMC1981}.

In the range of the attractive on-site interaction the structure of PDs of model (\ref{row:ham}) depends on the ratio $W/I$ only (cf. Fig \ref{rys:PSU0} for $U/I_0 = -1$). One can distinguish two ranges of the ratio $W/I$ in which the system exhibits substantially different behaviours:

(i)~\mbox{$0\leq W/I<1$}.
In Figs.~\ref{rys:PSU0}a,b, we present particular PDs for $W/I=0.5$.
For \mbox{$0<W/I<1$} and \mbox{$U<0$}, only the second-order  \mbox{SS--NO} transitions occur with increasing temperature.
If we analyze the system for fixed $n$, for this range of model parameters, the PS states do not occur and the obtained PDs have the same structure as those derived in \cite{RP1993,R1994,KRM2012}. The transition between homogeneous SS and NO phases taking place with increasing temperature is second order for arbitrary $\bar{\mu}$ and $n$,  and it decreases monotonically with increasing $|\bar{\mu}|/I_0$ and $|1-n|$.

(ii)~\mbox{$1<W/I$}.
A few particular PDs in this regime are presented in Figs.~\ref{rys:PSU0}c-f.
For $W/I=1$ the SS, the CO and the M phases are degenerate at $n=1$. For $W/I>1$,  three homogeneous phases (SS, CO, NO) appear. The SS--NO and CO--NO transitions are of the second order and these transition temperatures are decreasing function of $|\bar{\mu}|$ and $|1-n|$. The SS-CO transition is discontinuous for fixed $\bar{\mu}$, and thus the PS state SS-CO is stable in the definite range of $n$.
All transitions lines meet at a bicritical point $\mathbf{B}$.
With increasing $W$, the $\mathbf{B}$-point moves along the boundary between SS and NO phases toward larger $|\bar{\mu}|$ ($|1-n|$). This is due to the fact that in the VA the SS--NO transition is independent of $W$ \wyroz{(for fixed $n$)}. The region of the CO phase occurrence is extended, whereas the region of the SS phase stability is reduced by increasing the ratio $W/I$. The first-order SS-CO as well as the ``third-order'' SS-PS and PS-CO transition temperatures increase with $|\bar{\mu}|$ and $|1-n|$, respectively.

One should notice that for $W_{ij}$ restricted to nearest neighbours ($W_2=0$) the PS state is  strictly degenerated at $T=0$ with the M phase in the whole range of stability of both these states \cite{RP1996,PR1996}. This degeneration is removed at $T>0$, even for $W_2=0$ \wyroz{and the PS state occurs on PDs}. Repulsive $W_2>0$ between next-nearest neighbours destabilizes the PS state with respect to the M phase, whereas attractive $W_2<0$ extends the stability region of PS state and eliminates the M phase.

In the case of attractive $W<0$ (precisely for $W/I<-1$), the model can exhibit phase separation NO-NO (electron droplets state) at low temperatures \wyroz{\cite{KR2012,RP1996,PR1996,BT1993}} and for $U<0$ the PDs as a function of $n$ have the similar structure as those derived in \cite{BT1993}. We leave deeper analysis of this problem to future publications.

\subsection{Order parameters and thermodynamic properties}

\begin{figure}
    \centering
    \includegraphics[width=0.5\textwidth]{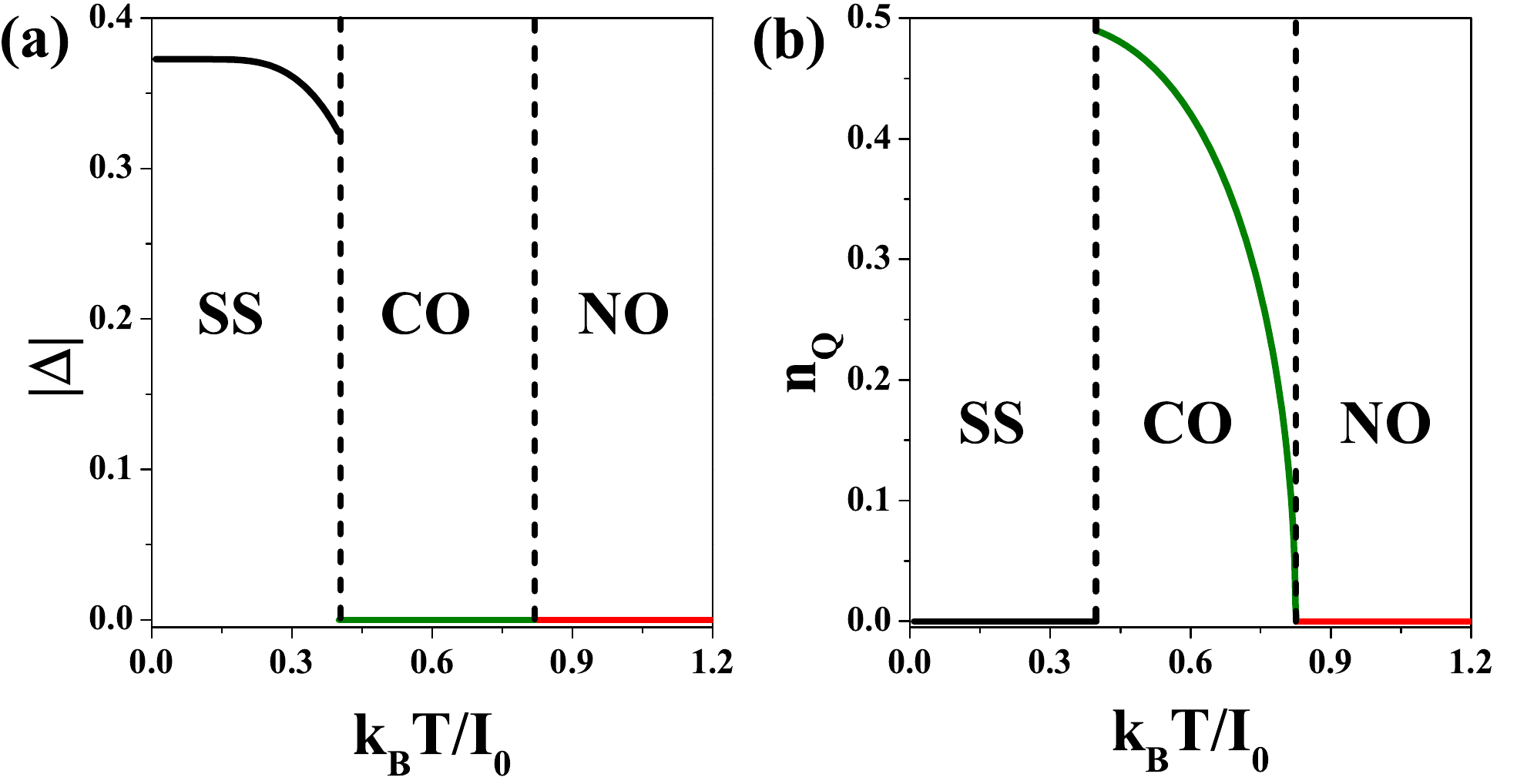}
\caption{Temperature dependencies of (a) superconducting order parameter $|\Delta|$ and (b) charge order parameter $n_Q$ for $W/I=2.0$, $U/I_0 = -1.0$ and $\bar{\mu}/I_0=-2.0$.}
\label{fig:order}       
\end{figure}

\begin{figure*}
  \centering
  \includegraphics[width=1.0\textwidth]{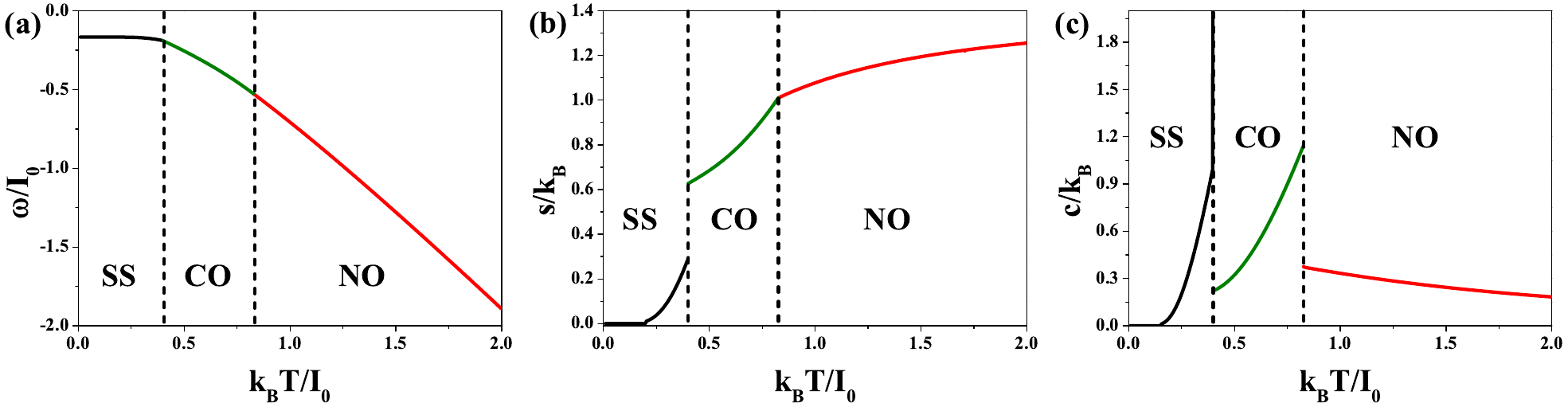}
\caption{Temperature dependencies of thermodynamics parameters: (a) the grand potential $\omega$, (b) the entropy $s$, and (c) the specific heat $c$ for $W/I=2.0$, $U/I_0 = -1.0$ and $\bar{\mu}/I_0=-2.0$.}
\label{fig:param}       
\end{figure*}

Let us focus now on the temperature dependencies of the order parameters and thermodynamic properties of the system at the sequence of transitions: SS $\rightarrow$ CO $\rightarrow$ NO for $W/I=2.0$, $U/I_0 = -1.0$ and $\bar{\mu}/I_0=-2.0$.

The temperature dependencies of the order parameters: $\Delta$ and $n_Q$ are presented in Fig.~\ref{fig:order}. \wyroz{It} is clearly seen that at the SS-CO transition (at $k_BT_{c1}/I_0 = 0.40$) the both order parameters change discontinuously. The CO-NO transition at $k_BT_{c2}/I_0 = 0.83$ is of the second order.

Calculating the grand potential per site
$\omega=-{1}{/(N\beta)}\ln{(\textrm{Tr}[\exp(-\beta\hat{H})])}$
within the VA one can obtain thermodynamic characteristics of the system for arbitrary temperature. The entropy $s$ and the specific heat $c$ can be derived as $s=-(\partial\omega/\partial T)$ and $c=-T (\partial^2\omega/\partial T^2)$. $\omega$, $s$, and $c$ as a function of temperature are shown in Fig.~\ref{fig:param}.
$s$ increases monotonically with increasing $T$.  It is discontinuous at $T_{c1}$ whereas it is continuous at $T_{c2}$.
One can notice that in the high-temperature limit \mbox{$s/k_B\rightarrow \ln(4) \approx 1.386$} (there are four possible configurations at each site).
The peak in $c(T)$ is associated with the first-order transition (at $T_{c1}$), while the \mbox{$\lambda$-point} behaviour is typical for the second-order transition (at $T_{c2}$).

\section{Conclusions and final remarks}

In this report, we have presented some particular PDs of the extended Hubbard model with pair hopping and intersite density-density interactions in the zero-bandwidth limit for the case of local attraction $U<0$.
One finds that the system considered can exhibit very interesting multicritical behaviours. Our investigations  show that, depending on the values of interaction parameters (the ratio $W/I$), the system can exhibit three homogeneous phases: superconducting, charge-ordered and nonordered. The SS-NO and CO-NO transitions are of the second order. The SS-CO transition is discontinuous (for fixed $\mu$), what leads to phase separation on the phase diagrams as a function of $n$ \wyroz{for $W/I>1$}. The homogeneous mixed phase (with nonzero both charge-ordered and superconducting order parameters) never occurs on PDs at $T>0$, at least in the absence of the next-nearest neighbours interactions. On the contrary the PS state: SS-CO is found to be stable in definite ranges of model parameters and temperatures.

Our results are exact in the limit of infinite dimensions, where the MFA treatment of intersite terms is rigourous one \wyroz{\cite{RP1993,R1994,KRM2012,KR2013,HB1977}}. In finite dimensions due to quantum fluctuations connected with the $I$ term, the regions of the ordered homogeneous phases occurrence are extended in comparison with the VA results \wyroz{\cite{KRM2012,KR2013}}.

\begin{acknowledgments}
The author is indebted to Professor Stanis\l{}aw Robaszkiewicz for very fruitful discussions during this work and careful reading of the manuscript.
The work has been financed by National Science Center (NCN) as a research project in the years 2011-2013, under Grant No. DEC-2011/01/N/ST3/00413.
We thank the
European Commision
and the Ministry of Science and Higher Education
(Poland) for the partial financial support from European Social Fund -- Operational Programme ``Human Capital'' -- POKL.04.01.01-00-133/09-00 -- ``\textit{Proinnowacyjne kszta\l{}cenie, kompetentna kadra, absolwenci przysz\l{}o\'sci}''
\wyroz{as well as the Foundation of Adam Mickiewicz University in
Pozna\'n for the support from its scholarship programme}.
\end{acknowledgments}

\end{document}